\documentclass[12pt,a4paper]{conference}
\usepackage{fancyhdr}
\usepackage{graphicx,amsmath,amssymb,cite}
\usepackage{multind}
\makeindex{author} \makeindex{subject}

\newcommand{\beq}{\begin{equation}}
\newcommand{\eeq}[1]{\label{#1}\end{equation}}
\newcommand{\eeqn}{\end{equation}}

\newcommand{\beqa}{\begin{eqnarray}}
\newcommand{\eeqa}[1]{\label{#1}\end{eqnarray}}
\newcommand{\eeqan}{\end{eqnarray}}

\let\bar=\overbar

\newcommand{\Dslash}{\not{\hbox{\kern-4pt $D$}}}
\newcommand{\dslash}{\not{\hbox{\kern-2pt $\del$}}}

\newcommand{\msb}{{\bar{\ssstyle M \kern -1pt S}}}

\begin{document}

\Chapter{Two-pion-exchange effects in $pp\to pp\pi^0$ reaction }
           {Two-pion-exchange effects in $pp\to pp\pi^0$ reaction}{ F. Myhrer}
\vspace{-6 cm}\includegraphics[width=6 cm]{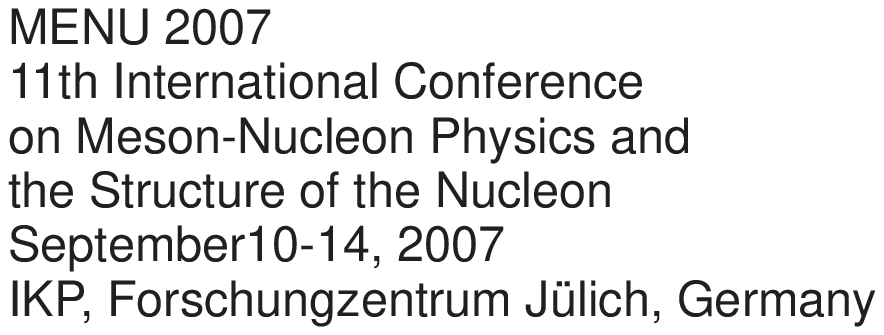}
\vspace{4 cm}

\addcontentsline{toc}{chapter}{{\it N. Author}} \label{authorStart}

\begin{raggedright}

{\it F. Myhrer }\index{author}{Author, N.}\\
Dept. of Physics and Astronomy \\
University of South Carolina \\
Columbia, SC 29208, USA 
\bigskip\bigskip

\end{raggedright}

\begin{center}
\textbf{Abstract}
\end{center}
We study the $pp\to pp\pi^0$ reaction near threshold based 
on heavy-baryon chiral perturbation theory. We 
show that the two-pion-exchange diagrams give much 
larger contribution than the one-pion-exchange diagram 
which is of lower chiral order in Weinberg's counting scheme. 
We also discuss the relation of our 
results to the momentum counting scheme.

\vspace*{10mm}


The near-threshold $pp$$\to$$pp\pi^0$ reaction 
has been attracting much theoretical attention,  
ever since experimental data of extremely high quality
became available~\cite{meyetal90}.
The heavy-baryon chiral perturbation theory 
(HB$\chi$PT) offers a possible systematic 
approach to the investigation of this reaction.
A motivation of this study may be stated
in reference to the generic
$NN$$\to$$NN \pi$  processes near threshold.
Although HB$\chi$PT presupposes the small size of its expansion
parameter  $Q/\Lambda_\chi$,
the pion-production reactions involve
somewhat large energy- and three-momentum transfers
even at threshold ($\vec{p}^{\; 2} \sim m_\pi m_N$).
Therefore, the application of HB$\chi$PT to 
the $NN$$\to$$NN \pi$  reactions
may involve some delicate aspects, 
but this also means that
these processes may serve as a good test case
for probing the limit of applicability of HB$\chi$PT.
Apart from this general issue, 
a specific aspect of
the $pp$$\to$$pp\pi^0$ reaction makes its study 
particularly interesting.
For most isospin channels, the $NN$$\to$$NN\pi$ 
amplitude near threshold
is dominated by the pion rescattering diagram where the
$\pi N$ scattering vertex is given by
the Weinberg-Tomozawa term,
which represents the lowest 
order contribution.
However, a quantitatively reliable 
description of  the $NN$$\to$$NN\pi$ reactions
obviously requires detailed examinations
of the corrections to this dominant amplitude.
Meanwhile, since the Weinberg-Tomozawa vertex does not
contribute to the pion-nucleon rescattering diagram
for $pp$$\to$$pp\pi^0$,
this reaction is particularly sensitive to higher 
order contributions
and hence its study is expected to provide 
valuable information to guide us
in formulating a quantitative description of  all the
$NN$$\to$$NN\pi$ reactions.

\vspace{3mm}  

At threshold the strong (and Coulomb) $pp$ initial and final state 
interactions 
have to be considered in these reactions. 
In our DWBA evaluation 
the nuclear transition operators 
are derived using HB$\chi$PT, 
whereas the initial and final state 
nuclear wave functions are calculated in 
the standard nuclear physics approach (SNPA). 
A serious problem encountered in performing such a 
{\it hybrid} $\chi$PT (for short called EFT$^*$ below) 
calculation
of pion production is that
the calculation involves uncomfortably high momentum components 
which are present in the nuclear wave functions.  
The occurrence of these high-momentum components 
goes against the tenet of $\chi$PT,
which presupposes the existence of 
a momentum cutoff scale,  
$\Lambda_\chi \simeq $ 1 GeV/c.
The high momenta arise from two sources. 
The first source is the 
large momentum components contained 
in the distorted initial and final wave functions
generated by a so-called high-precision  
phenomenological NN-potential, $V_{NN}$; 
see e.g. Ref.~\cite{pmmmk96}(b). 
The second source of the high momentum behavior 
originates from 
higher powers of momentum terms which 
appear in the transition amplitudes   
generated by higher $\chi$PT diagrams~\cite{dkms99,hk02}. 
Below we will focus our discussion on 
these transition amplitudes.

\vspace{2mm}

In order to eliminate from the $NN$ wave functions the 
high-momentum components 
that lie above the
original cutoff scale of $\chi$PT,
a suitably parameterized cutoff factor 
is introduced (we make 
certain the observables are independent of this cut-off). 
This is admittedly an operational remedy,
the foundation of which needs to be examined
from a formal point of view. 
It is also informative and of practical value  
to examine the use of the 
``low-momentum regime NN potential",
$V_{low-k}$~\cite{meissner99,birse99,kuo01}. 
$V_{low-k}$ is derived from 
$V_{NN}$
by integrating out the high-momentum components
contained in $V_{NN}$. 
Since $V_{low-k}$ by construction is free from
high-momentum components,
its use in an EFT* calculations for pion production
should alleviate the ``high momentum problem" 
that plagued the past DWBA calculations. From a purist's 
point of view
this may not be a totally satisfactory approach 
but we believe that this ``pragmatic" method 
still has its merits. 
We remark that, as is well known, $V_{low-k}$s 
generated from any realistic phenomenological potential 
lead to practically equivalent half-off-shell NN K-matrices 
and hence the same NN wave function.


\vspace{2mm}

We derived the 
TPE transition amplitude operators~\cite{dkms99} 
using WeinbergÕs chiral counting scheme with 
expansion parameter 
$\epsilon \simeq m_\pi / m_N \simeq 0.15$. 
We isolated the   
high-momentum components of these amplitudes 
using an asymptotic expansion, see 
Refs.~\cite{kkms06,ksmk07a}.
Hanhart and Kaiser~\cite{hk02}
used the momentum counting scheme 
(MCS)~\cite{cfmv96,hmk00}  to 
evaluate TPE diagrams for the reaction $NN$$\rightarrow$$NN\pi$. 
The MCS has the expansion parameter
$\tilde{\epsilon} \simeq (m_\pi/m_N)^{1/2} \simeq 0.39$.
Unlike Weinberg's chiral counting a subtlety 
in MCS is that loop diagrams of a given 
order $\nu$ in $\tilde{\epsilon}$ not only 
contains a contribution of order $\nu$ 
(the ``leading'' part) but, in principle, 
can also involve contributions of higher 
order in $\tilde{\epsilon}$ (``sub-leading'' parts). 
Hanhart and Kaiser evaluated 
the ``leading'' part  of the 
lowest MCS-order TPE diagrams
and showed that the ``leading'' parts of the 
two-pion exchange diagrams, when summed up,
cancel among themselves, see also Ref.~\cite{Lensky05}. 
We~\cite{sm99} identified the ``leading'' part  
of our TPE operators~\cite{dkms99} 
and confirmed this cancellation~\cite{kkms06,ksmk07a}. 
As will be discussed below, we have found however that 
the remainder, or the ``sub-leading'' parts, 
of the TPE amplitudes can be at 
least as large as the one-pion 
rescattering amplitude~\cite{kkms06,ksmk07a}. 
We consider it important to re-examine the behavior 
of MCSÕs ``sub-leading'' parts of these 
TPE diagrams in order to see 
whether they can still be as large as indicated 
by the phenomenological SNPA success of the 
Lee-Riska heavy-meson exchange mechanism. 
In a forthcoming publication~\cite{ksmk07b} we 
will also consider other chiral correction amplitudes
including the contributions from counter-terms needed
to regulate the UV behavior of 
the TPE loop diagrams. 
\begin{figure}[hb]
\begin{center} 
\includegraphics[height=7.0 cm, width=9.0 cm]{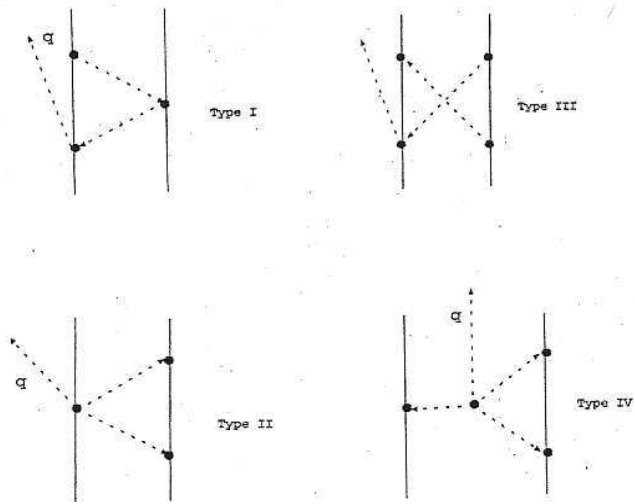}
\caption{The two-pion-exchange loop-diagrams discussed in the text.}
\label{fig:TPE}
\end{center}
\end{figure}

\vspace{2mm}

The one-pion loop $pp\to pp\pi^0$ 
transition operators were evaluated analytically by  
Ref.~\cite{dkms99} using HB$\chi$PT. 
When these operators are sandwiched with 
phenomenological determined 
distorted initial and final $NN$ wave 
functions, we find that 
the momentum integrals convergence 
very slowly \cite{kkms06,sm99}. 
This slow convergence can easily be understood 
when we adopt the  threshold 
fixed kinematics approximation ({\it FKA}) 
to evaluate the amplitudes.  
We impose the {\it FKA} on the analytic expressions for the 
transition operators of the different  
TPE diagrams are given in  
Ref.~\cite{dkms99}   and make an 
asymptotic expansion in the two-nucleon 
momentum transfer $ \vec{k} $, i.e. 
$ |\vec{k} | = 
|\vec{p}-\vec{p}^{\; \prime}| \to \infty $. 
The transition operator matrix $T$ of 
the TPE diagrams 
is of the form 
\begin{eqnarray}
T &=& \left( \frac{g_A}{f_\pi} \right) 
\left( \vec{\Sigma}\cdot \vec{k} \right) 
t(p,p^\prime , x) 
\label{eq:T}
\end{eqnarray} 
where 
$x=\hat{p}\cdot\hat{p}^\prime$. 
The generic {\it asymptotic} 
behavior for 
$ t(p,p^\prime , x)$ is \cite{sm99}:
\begin{eqnarray}
t(p,p^\prime , x) &\stackrel{k\to\infty}{\sim} \;& 
\;t_1\! \left( g_A/(4f_\pi^2)\right)^2 
| \vec{k} | + t_2 \; 
{\rm ln}[\Lambda^2/ |\vec{k} |^2 ] 
+ t_3 + \delta t(p,p^\prime , x) \; , 
\label{eq:tasympt} 
\end{eqnarray} 
where $t_3$ is asymptotically $k$-independent,
and $\delta t(p,p^\prime , x) $ is  ${\cal O}(k^{-1})$.
For each of Types I $\sim$ IV,
analytic expressions for $t_i$'s ($i=1, 2, 3$)
can be extracted~\cite{sm99} from
the amplitudes $T$ given in Ref.~\cite{dkms99}.
The first term with $t_1$ in eq.(\ref{eq:tasympt}) is
the ``leading'' term in MCS discussed by HK~\cite{hk02},
whereas the remaining terms,
which we refer to as the ``sub-leading" terms,
were not considered by HK.
%
\begin{center}
 \parbox[t]{5.3in}{ Table 1:
For the four types of TPE diagrams,
K= I, II, III and IV,
the second row gives the value of $t_1$ defined in eq.(\ref{eq:tasympt}),
and the third row gives the ratio
$R_K = T_K /T_{Resc}$, where $T_K$ is
the plane-wave matrix element of
$T$ in eq.(\ref{eq:T}) for Type K,
and $T_{Resc}$ is the lowest chiral order 
one-pion-exchange rescattering (Resc)  amplitude.
The last row gives
$R_K^{\,\,\star} = T_K^{\,\,\star} /T_{Resc}$,
where $T_K^{\,\,\star}$ is
the plane-wave matrix element of
$T$ in eq.(\ref{eq:T}) with the $t_1$ term in eq.(\ref{eq:tasympt})
subtracted.}
\end{center}
$$
\begin{array}{|l| |r| r|r|r|r|}
\hline
{\rm Type\; of\; diagrams:  K=} & {\rm I}&{\rm II}&{\rm III}&{\rm IV}
& {\rm Sum}  \\
\hline
 (t_1)_K \;  \propto \; \; \; & 0 & -1  & -1/2 & 3/2 & 0  \\
\hline
 R_K  & -.70& -6.54& -6.60 & 9.19  & -4.65 \\
\hline
R^{\,\star}_K& -.70& -0.82& -3.73& 0.60 & -4.65 \\
\hline
\end{array}
$$


In Table 1, the second row shows that the ``leading'' parts 
in the MCS $\propto t_1$ of the TPE diagrams I $\sim$ IV 
sum to zero. 
This confirms the finding of 
Hanhart and Kaiser \cite{hk02}. 
In the third 
row marked $R_K$, and using {\it FKA}, we give 
the values of the ratio of the transition amplitude 
in Ref.~\cite{dkms99}
to the rescattering amplitude in 
the plane wave approximation for 
the four TPE amplitudes. 
We note that the sum of the TPE amplitudes are non-zero. 
The magnitude of the sum reflects the size of 
the ``sub-leading'' parts of the TPE amplitudes. 
In the fourth row we have \underline{removed}
the ``leading'' term, $t_1$ from 
$t(p,p^\prime,x)$ in Eq.(\ref{eq:T}) and then evaluated 
the similar ratio $R_K^{\,\,\star}$. 
We observe that the sum of the four
modified TPE amplitudes (II, III and IV) is larger 
than the rescattering amplitude~\cite{kkms06}. 
This can also be inferred from the $R_K$ 
row of the table where the ratios of the 
amplitudes II:III:IV are about  \\ 
-2:-2:3, 
whereas the ratios of 
the corresponding amplitudesÕ 
``leading'' terms  are -2:-1:3. 
However, the evaluations of the different 
diagrams appear to confirm 
that the magnitudes of the different diagrams 
follow the 
momentum counting rule as indicated in 
Table 11 of Ref.~\cite{hanhart04}. 

\vspace{2mm}

Next we investigate the behavior of the TPE diagrams 
as we go beyond the plane-wave approximation by 
using distorted waves (DW) for the initial- and final-state 
NN wave functions. For formal consistency we 
should use the NN potential recently 
derived from HB$\chi$PT~\cite{bira,evgeny}, 
but as discussed earlier
we adopt here EFT$^*$. 
As argued earlier, in order to
stay close to the spirit of HB$\chi$PT, 
we introduce a Gaussian momentum regulator, 
exp($-p^2/\Lambda_G^2$), in the initial and 
final distorted wave integrals, suppressing thereby 
the high momentum components of the  
phenomenological NN potentials, e.g. the 
Bonn~\cite{Bonn} or  
Nijmegen potential~\cite{nijmegen}. 
We require that $\Lambda_G$ be 
larger than the characteristic momentum scale of 
the $NN \to NN\pi$ reactions, 
$| \vec{p} | \sim \sqrt{m_\pi m_N} \simeq 360$ MeV/c, 
but it should not exceed the chiral scale 
$\Lambda_\chi \sim 1$ GeV/c. 
Part of the work to implement the idea of utilizing 
a low-momentum regime NN potential 
has been published~\cite{kdkms06} where use was made of  
$V_{low-k}$.  
One issue regarding the use of Stony Brook's $V_{low-k}$ 
in the present context 
is that it is derived with a rather low value
of the cutoff parameter, $p_{max} = 2$ fm$^{-1}$, 
which is very close to the threshold momentum for the 
$NN$$\to$$NN\pi$ reactions.  
We therefore have extended the $V_{low-k}$ potential
to cases where the momentum cut-off $p_{max}$ 
is allowed to be larger than the original Stony Brook value
up to 5 fm$^{-1}$ (this value  corresponds to 
the chiral scale, $\Lambda_\chi \simeq 1$ GeV/c$^2$). 
This extended $V_{low-k}$ has been used in our 
recent study of the two-pion-exchange (TPE) amplitudes
for the $pp$$\to$$pp\pi^0$ reaction~\cite{ksmk07a}. 

\vspace{2mm}

To evaluate the $pp\to pp\pi^0$ reaction 
at threshold using 
HB$\chi$PT, the impulse approximation (I.A.) 
and the 
one-pion-exchange (Resc) diagrams 
are the 
lowest order amplitudes (diagrams) according to 
Weinberg counting. 
However, the typical momentum for this 
reaction at threshold is 
$p \sim \sqrt{m_\pi m_N}$, which implies 
that for this reaction we have to 
take Weinberg chiral counting with a 
grain of salt. 
It was shown early on that 
in HB$\chi$PT the I.A. and Resc amplitudes interfere 
destructively resulting in a 
very small cross sections~\cite{pmmmk96,cfmv96}. 
We therefore expect sizeable contributions 
to this reaction from the TPE diagrams. 
In Table 2 we show examples, taken from Ref.~\cite{ksmk07a},  
of DWBA evaluations for a typical energy 
$T_{lab} = 281$ MeV. 
Since the $t_1$ terms 
add to zero and to improve the numerical convergence, 
we drop the $t_1$ terms 
in our calculations as was done in the fourth row of Table 1. 
Thus, in Eq.(\ref{eq:T}), we use $t^\star(p,p',x)$
instead of $t(p,p',x)$,
where $t^\star(p,p',x)$ is obtained from $t(p,p',x)$
by suppressing the $t_1$ term in Eq.(\ref{eq:tasympt}).
The partial-wave projected form of
$t^\star(p,p^\prime,x)$ in a DWBA calculation
is written as:
\begin{eqnarray}
J = -\left(\frac{m_Nm_\pi}{8\pi}\right)
\int_0^\infty p^2 {\rm d}p\; p^{\prime\; 2}
{\rm d}p^\prime
\int_{-1}^1 {\rm d}x\; \psi_{^1\!S_0}(p^\prime )\;
t^\star(p,p^\prime,x) \; (p-p^\prime x) \psi_{^3\!P_0}(p)
\label{eq:J}
\end{eqnarray}
In Table 2 we show the values of 
the $J$ amplitudes for each TPE diagram.
\vspace{2mm}
\begin{center}
 \parbox[t]{5.3in}{ Table 2:
The values of $J$, Eq.(\ref{eq:J}), corresponding to
the TPE diagrams of Types I $\sim$ IV,
evaluated in a DWBA calculation for $T_{lab}= 281$ MeV.
The column labeled ``Sum" gives
the combined contributions of Types I $\sim$ IV,
and the last column gives the value of $J$ for 1$\pi$-Resc.
For the Nijm93 potential case,
the results for three different choices of  
$\Lambda_G$ (in MeV/c) 
are shown.
For the case with V$_{low-k}$,
CD-4 (CD-5) represents V$_{low-k}$
generated from the CD-Bonn potential
with a momentum cut-off
$\Lambda_{low-k}$ = 4\,fm$^{-1}$ (5\,fm$^{-1}$).
The last row gives the results obtained 
in plane-wave approximation.
}
\end{center}
$$
\begin{array}{|l||r|r|r|r|c||r| }
\hline
    &\!\! {\rm I}&{\rm II}\; \; &{\rm III}\; \; &{\rm IV} \; \; &
\;\:\:\;{\rm Sum} \:\:\; & 1\pi\!\!-\!\!{\rm Resc} \\
\hline
V_{\rm Nijm}\,:\Lambda_G=600 & -0.12 & -0.12 &  -0.57
& 0.07 & -0.74 &0.20 \\
\hline
V_{\rm Nijm}\,:\Lambda_G=700 & -0.12 & -0.11 &  -0.57
& 0.06 & -0.74 & 0.21\\
\hline
V_{\rm Nijm}\,:\Lambda_G=800 & -0.12 & -0.11 &  -0.55
& 0.04 & -0.74 & 0.22 \\
\hline
V_{low-k}  \; \; (CD\!-\!4) & -0.12  & -0.09&  -0.46
& 0.03 & -0.65 & 0.23  \\
\hline
V_{low-k}  \; \; (CD\!-\!5) & -0.09  & -0.06&  -0.30
& -0.01 & -0.46 & 0.22 \\
\hline \hline
{\rm Plane\!-\!waves} & -0.06   & -0.07 &  -0.30
& 0.05& -0.37 &  0.080 \\
\hline
\end{array}
$$

\vspace{3mm}

Table 2 shows that the DW amplitudes from the TPE diagrams 
are only roughly of the order of the 
one-pion rescattering amplitude tabulated 
in the last column. Evidently, 
when we compare to the plane wave amplitudes, the DWBA 
treatment does affect the relative magnitudes of the 
various diagrams differently. 
In the last row we show the plane wave amplitudes 
and we note that the ``sub-leading'' part 
of diagram III is 
a factor three or more larger than the other amplitudes.
The MCS indicates that the ``sub-leading'' parts of 
the TPE amplitudes should be of the same order as 
the one-pion rescattering amplitude. 
Furthermore, the expansion parameter in MCS,
$\tilde{\epsilon} \simeq 0.4$, ideally 
speaking should be the ratio of the 
amplitudes from the different orders in the MCS. 
When comparing the results in Table 1 and Table 2, we 
find 
that the ``leading'' part of 
diagrams II and IV are almost an 
order of magnitude larger than 
their ``sub-leading'' parts.
Clearly, as seen in Table 2, we find, as expected in the MCS, that 
the amplitudes from the ``sub-leading'' parts of 
diagrams II, IV and from the 
one-pion rescattering diagram are 
about the same magnitude. 
However, diagram III (the ``cross-box'' diagram) 
appears 
not follow the expected MCS behavior. 
The plane wave amplitude for the ``sub-leading'' part of 
diagram III is less than 50\% than the ``leading'' 
part of diagram III. 
Moreover, the 
amplitude of the ``sub-leading'' part of 
diagram III is more than a factor $\tilde{\epsilon}^{-1}$ 
larger than what is expected in the MCS. 
One explanation could be that we evaluate diagram III using 
HB$\chi$PT's heavy nucleon propagators 
and not the nucleon propagator which is 
advocated for the MCS~\cite{hanhart04}. 
This issue will be resolved in the near future. 
A final note, the lowest- and next-chiral-order 
(diagram VII in Ref.~\cite{dkms99}) 
one-pion-exchange diagrams were found to be same 
order of magnitude as expected in MCS. 

\vspace{2mm} 

We have demonstrated~\cite{ksmk07a} that, as expected, 
the two-pion-exchange 
loop diagrams give very large contributions to the 
$pp\to pp\pi^0$ reaction at threshold, 
and that these diagrams will give important 
contributions to other $NN\to NN\pi$ reaction channels.

\section*{Acknowledgments}

This research has been done in a close 
collaboration with Drs. Y. Kim, K. Kubodera and T. Sato. 
This work is supported in part by the National Science Foundation, 
Grant No. PHY-0457014.



\end{document}